\documentclass[12pt,epsbox]{article}
\usepackage[dvips]{graphicx}
\usepackage{amsmath}

\baselineskip=\normalbaselineskip

\newcommand{\resection}[1]
 {\setcounter{equation}{0}\section{\large{#1}}}

\newcommand{\bel}[1]{\begin{equation}\label{#1}}
\newcommand{\bal}[1]{\begin{eqnarray}\label{#1}}

\newcommand{\be}{\begin{equation}}
\newcommand{\ee}{\end{equation}}
\newcommand{\ba}{\begin{eqnarray}}
\newcommand{\ea}{\end{eqnarray}}
\newcommand{\nn}{\nonumber \\}

\newcommand{\bR}{{\bf R}}

\newcommand{\hg}{\widehat{g}}

\newcommand{\eq}[1]{(\ref{#1})}

\newcommand{\bG}{\mbox{\boldmath $G$}}
\renewcommand{\bR}{\mbox{\boldmath $R$}}
\newcommand{\bS}{\mbox{\boldmath $S$}}

\newcommand{\hR}{\widehat{R}}

\begin{document}
\setcounter{page}{0}
\begin{flushright}
\parbox{40mm}{%
KUNS-1782 \\
YITP-02-30 \\
{\tt hep-th/0204257} \\
April 2002}

\end{flushright}

\vfill

\begin{center}
{\large{\bf 
Comment on Field Redefinitions\\ 
in the AdS/CFT Correspondence
}}
\end{center}

\vfill

\begin{center}
{\sc Masafumi Fukuma}
\footnote{E-mail: {\tt fukuma@gauge.scphys.kyoto-u.ac.jp}} and 
{\sc So Matsuura}
\footnote{E-mail: {\tt matsu@yukawa.kyoto-u.ac.jp}} \\[2em]
$^1${\sl Department of Physics, Kyoto University, Kyoto 606-8502, Japan} \\
$^2${\sl Yukawa Institute for Theoretical Physics, 
      Kyoto University, Kyoto 606-8502, Japan } \\

\end{center}

\vfill
\begin{center}
ABSTRACT
\end{center}

\begin{quote}

\small{%
We carry out field redefinitions in ten-dimensional Type IIB
supergravity 
and show that they do not give rise to any physical corrections 
to the holographic renormalization group structure 
in the AdS/CFT correspondence. 
We in particular show that the holographic Weyl anomaly of 
the ${\mathcal{N}}=4$ $SU(N)$ super Yang-Mills theory 
does not change under the field redefinition 
of the ten-dimensional metric of the form 
$\bG_{MN}\rightarrow\bG_{MN}+\alpha\bR\bG_{MN}+\beta\bR_{MN}$. 
These results are consistent with the fact  
that classical supergravity represents the on-shell 
structure of massless modes of superstrings, 
which should not change under redefinitions of fields.
}
\end{quote}
\vfill

\renewcommand{\thefootnote}{\arabic{footnote}}
\setcounter{footnote}{0}
\addtocounter{page}{1}
\resection{Introduction}

The AdS/CFT correspondence \cite{ads/cft}--\cite{review}
asserts that  
the classical theory of $(d+1)$-dimensional gravity in an AdS background 
is related to a $d$-dimensional CFT at the 
boundary of the AdS geometry. 
More precisely, we can regard 
an on-shell field in the gravity theory as the source 
coupled to a scaling operator in the CFT at the boundary.
Among many applications of the AdS/CFT correspondence, 
the holographic renormalization group (RG) 
\cite{SW}--\cite{HS} is 
one of the most important. 
In the holographic RG, we regard 
the radial coordinate of the $(d+1)$-dimensional manifold 
as a scaling parameter of the corresponding boundary field theory.
Using this scheme, we can describe many aspects of 
the RG structure of the $d$-dimensional boundary field theory 
using the $(d+1)$-dimensional classical gravity theory. 
For example, we can derive the Callan-Symanzik equation 
of the corresponding $d$-dimensional boundary field theory 
from the Hamilton-Jacobi equation 
of the $(d+1)$-dimensional gravity theory \cite{dVV}, 
which gives us a systematic formulation of the holographic RG 
(see also \cite{FMS}\cite{FS}\cite{NOO}). 

There have been numerous quantitative studies to check the validity 
of the AdS/CFT 
correspondence and the holographic renormalization group. 
Among such studies are calculations of the chiral anomaly \cite{FMMR} 
and the Weyl anomaly \cite{HS} of the four-dimensional 
${\mathcal{N}}=4$ $SU(N)$ super Yang-Mills theory (SYM${}_4$), 
which is believed to be realized on the boundary of AdS${}_5$ 
after the ten-dimensional spacetime is factorized as AdS${}_5\times S^5$.  
Both calculations were carried out purely on the basis of five-dimensional 
supergravity theory and correctly reproduce the field-theoretical 
results in the large $N$ limit.

In this article, as another study to check the validity of the 
AdS/CFT correspondence, we show that the holographic RG structure 
does not undergo any physical corrections 
under field redefinitions of ten-dimensional supergravity. 
The AdS/CFT correspondence should have this property, 
since classical supergravity represents the on-shell 
structure of massless modes of superstrings, 
and the on-shell amplitudes (more precisely, the residues of 
one-particle poles of correlation functions) should be invariant under 
redefinitions of fields \cite{KOS} (see also \cite{GW} 
for discussions in the context of string theory).%
\footnote{See also \cite{LM} for recent 
discussion about scheme independence in the renormalization group structure.}  

It is easy to demonstrate the invariance of the holographic RG structure 
for point-transformations of scalar fields in supergravity, 
\ba
 \phi^I\rightarrow \phi^{\prime\,I}=f^I(\phi),
\ea 
because the superpotential $W(\phi)$ transforms as 
a scalar over the space parametrized by $\phi^I$: 
$W(\phi)\rightarrow W'(\phi)=W(f(\phi))$, 
so that the beta function of the boundary field theory 
transforms as a vector field over such space 
\cite{dVV}\cite{FMS}:\footnote{%
$L_{IJ}(\phi)$ is the metric on the space $\{\phi^I\}$ and  
$c(\phi)=\bigl(-W(\phi)\bigr)^{-(d-1)}$ can be 
identified with the $c$-function.}
\ba
 \beta^I(\phi)\Bigl(=-\frac{2(d-1)}{W(\phi)}\,L^{IJ}(\phi)\,
  \frac{\partial}{\partial\phi^J}W(\phi)\Bigr)
  \rightarrow \beta^{\,\prime\,I}(\phi)=
  \frac{\partial \phi^I}{\partial f^J}\,\beta^J(f(\phi)).
\ea

Similar arguments can be applied to field redefinitions 
that include derivatives of fields, 
such as the redefinition of the ten-dimensional metric 
of the form $\bG_{MN}\rightarrow 
\bG_{MN}+\alpha\bR\bG_{MN}+\beta\bR_{MN}$. 
In this case, however, 
the resulting gravity action obtained after such redefinitions 
possesses higher-order derivative terms. 
Thus, after the compactification on $S^5$, 
one needs to treat the five-dimensional gravity theory 
with curvature squared terms. 

The structure of the holographic RG for higher-derivative gravity 
was investigated generally 
in Refs.\ \cite{NO}\cite{BGN}\cite{FMS2}\cite{FM}, 
where it is shown that if the five-dimensional gravity action is 
given by\footnote{%
The cosmological constant is parametrized in such a way 
that the classical solution can have an AdS spacetime 
with radius $L$.} 
\ba
 \bS_5 = {1\over 2\kappa_{5}^2}\int d^5x \sqrt{-\hg} 
 \left[{12\over L^2}-{80a+16b+8c\over L^4}
 +\hR+a\hR^2+b\hR_{\mu\nu}^2+c\hR_{\mu\nu\rho\sigma}^2\right], 
 \label{weyl-1}
\ea
then the Weyl anomaly of the corresponding boundary CFT is 
\begin{align} 
 \langle T_i^i \rangle &= 
 {2L^3\over2\kappa_5^2}\biggl[
 \Bigl(1+\frac{8(5a+b+c)}{L^2}\Bigr)
 \Bigl(-{1\over24}R^2+{1\over8} R_{ij}^2\Bigr)
 +{c\over 2L^2}R_{ijkl}^2
 \biggr].
 \label{weyl-2}
\end{align}
From this, it is seen that if $c$ vanishes, then it may be possible 
to absorb the change $(1+8(5a+b)/L^2)$ into the five-dimensional 
Newton constant $2\kappa_5^2$. 
In fact, for the field redefinitions of the form 
$\bG_{MN}\rightarrow\bG_{MN}+\alpha\bR\bG_{MN}+\beta\bR_{MN}$, 
no terms including the Riemann tensor $\bR_{KLMN}$ are induced, 
so that we only have to consider the case where $c=0$. 
Furthermore, as we show in the following sections, 
the field equation in ten dimensions changes 
the radius of $S^5$ exactly in such a way 
that the change of the five-dimensional Newton constant, 
$2\kappa_5^2=2\kappa_{10}^2/{\rm volume}(S^5)$, 
cancels the factor $(1+8(5a+b)/L^2)$ together with 
the contribution from the Ramond-Ramond terms.

In \S 2, we derive the ten-dimensional Type IIB supergravity action 
that is obtained through the field redefinition, 
and then we discuss its AdS$_5\times S^5$ solution.
In \S3, after explaining how to determine the five-dimensional 
gravity action when the geometry is compactified on $S^5$, 
we calculate the holographic Weyl anomaly for 
${\mathcal{N}}=4$ $SU(N)$ SYM$_4$ 
and show that the result is exactly the same with that for 
the original anomaly 
before the field redefinition. 
Section 4 is devoted to conclusions.

\resection{Field Redefinition of Type IIB Supergravity and 
the AdS$_5\times$ ${\mathbf S}^5$ Solutions}

In this section, 
we consider a field redefinition in the ten-dimensional 
type IIB supergravity theory.
We first give the usual IIB supergravity action and 
its AdS$_5\times S^5$ solution.
We then carry out a field redefinition of the ten-dimensional metric 
and derive the corresponding action with its AdS$_5\times S^5$ solution.

We start with the bosonic part of 
the ten-dimensional Type IIB supergravity action 
given by\footnote{%
The coefficient of $|F_5|^2$ is chosen to be $(-1/4)$, 
which is one half of the canonical value $(-1/2)$. 
This is necessary for the action to be invariant under $T$-duality 
transformations (see, {\em e.g.}, \cite{FOT}).}
\ba
\bS_{10}=\frac{1}{2\kappa_{10}^2}\int d^{10}X \sqrt{-\bG}
\left[e^{-2\phi}\left(\bR+4\left|d\phi\right|^2\right)
-{1\over4}\left|F_5\right|^2\right].
\label{IIB-1}
\ea
Here $\phi$ and $F_5$ are the dilaton and the self-dual 
Ramond-Ramond 5-form field strength, respectively, 
and we have set other fields of Type IIB supergravity to zero. 
In this equation, we have used the definitions 
\ba
 \left|d\phi\right|^2
 \equiv \bG^{MN}\,\partial_M\phi\,\partial_N\phi,\quad 
 \left|F_5\right|^2\equiv\frac{1}{5!}\,\bG^{M_1N_1}\cdots\bG^{M_5N_5}\,
 (F_5)_{M_1\cdots M_5}(F_5)_{N_1\cdots N_5}.
\ea 
The self-duality of $F_5$ is imposed on the field equations 
(not in the action) as a constraint.

In the context of the AdS$_5$/CFT$_4$ correspondence, 
we are interested in an AdS$_5\times S^5$ solution 
that is realized 
as the near horizon limit of the black 3-brane solution 
\cite{p-brane}:
\begin{gather}
 ds^2 = \frac{l_0^2}{r^2}\,dr^2
        +\frac{r^2}{l_0^2}\,\eta_{ij}\,dx^idx^j
        +l_0^2 \,d\Omega_5^2, \nn
 (F_5)_{r0123}= -{4\over g_s}\,{r^3\over l_0^4}, \quad
 (F_5)_{y^1\cdots y^5}={4\over g_s}\,l_0^4, \nn
 e^{\phi}=g_s. 
  \label{sln-1}
\end{gather}
Here, $d\Omega_5^2=(\delta_{ab}+y_ay_b/(1-y^2))dy^a dy^b$ 
$(-1\!\leq\!y^a\!\leq\!1,~a,b\!=1,\cdots,5)$ is 
the metric of the unit five-sphere and $i,j\in\{0,1,2,3\}$.
In this case,
the AdS$_5$ and $S^5$ have the same radius, $l_0$, 
whose value is determined by the D3-brane charge as 
\ba
 l_0=(4\pi g_s N)^{1/4}, 
\label{radius}
\ea
where $N$ is the number of the coincident D3-branes, and we have set 
the string length $\sqrt{\alpha'}$ to $1$.

As discussed in the Introduction, we can make 
an arbitrary field redefinition $\delta \bG_{MN}=X_{MN}$ 
without changing the content of the Type IIB supergravity theory 
\cite{GW}. 
Now we make an infinitesimal change of the metric as 
\ba
 \bG_{MN} \to \bG_{MN}' \equiv  \bG_{MN}+\alpha \bR\bG_{MN}+\beta\bR_{MN} 
 \quad \bigl(F'_5\equiv F_5,\quad \phi'\equiv \phi\bigr).
 \label{redef}
\ea
Then the new gravity action is obtained as 
\begin{align}
 \widetilde{\bS}_{10}[\bG_{MN}] &\equiv \bS_{10}[\bG_{MN}'] \nn
 &=\bS_{10}[\bG_{MN}+\alpha \bR\bG_{MN}+\beta\bR_{MN}],
\end{align}
which is expressed explicitly as 
\begin{align}
 \widetilde{\bS}_{10} = \frac{1}{2\kappa_{10}^2}\int d^{10}X \sqrt{-\bG}
 \Biggl\{
  &e^{-2\phi}\biggl[\bR+4\left|d\phi\right|^2
  +a\bR^2 +b\bR_{MN}^2 \nn
  &\qquad\quad+a\bR\left|d\phi\right|^2
  +b\,\bR^{MN}\partial_M\phi\,\partial_N\phi\biggr] \nn
  &-{1\over4}\left|F_5\right|^2
  +{b\over8}\,\bR\left|F_5\right|^2
  -{b\over4}\,{1\over 4!}\,\bR_{MN}(F_5)^{MPQRS}(F_5)^N_{\ PQRS}
 \Biggr\}.
 \label{IIB-2}
\end{align}
Here $a$ and $b$ are defined as
\ba
 a=4\alpha+{1\over2}\beta,\quad
 b=-\beta.
\ea

Since $\bG'_{MN}$ and $F_5$ can be expressed as in Eq.\ \eq{sln-1}, 
\ba
 &&ds^{\prime 2} =\bG'_{MN}\,dX^M dX^N= \frac{l_0^2}{r^2}\,dr^2
        +\frac{r^2}{l_0^2}\,\eta_{ij}\,dx^idx^j
        +l_0^2\, d\Omega_5^2, \nn
 &&(F_5)_{r0123}= -{4\over g_s}\,{r^3\over l_0^4}, \quad
 (F_5)_{y^1\cdots y^5}={4\over g_s}\,l_0^4, \quad
 e^{\phi}=g_s, 
\ea
we can easily construct an AdS$_5\times S^5$ 
solution for the action (\ref{IIB-2}):
\ba
 ds^2&=&\bG_{MN}\,dX^M dX^N
  =\Bigl(\bG'_{MN}-\alpha \bR'\bG'_{MN}
   -\beta\bR'_{MN}\Bigr)dX^MdX^N\nn
 &=&\left(1-{8b\over l^2}\right){l^2\over r^2}\,dr^2
 +{r^2\over l^2}\,\eta_{ij}\,dx^idx^j
 +l^2 d\Omega_5^2, \nonumber
\ea
\begin{gather}
 (F_5)_{r0123} = {4\over g_s}\left(1+{8b\over l^2}\right){r^3\over l^4},
 \quad 
 (F_5)_{y^1\cdots y^5} = 
 {4\over g_s}\left(1-{8b\over l^2}\right)l^4,\quad
 e^{\phi} = g_s. \label{sln-2}
\end{gather}
Here we have used the fact that with the solution 
(\ref{sln-1}), the Ricci tenser becomes 
\ba
 \bR_{\mu\nu}=-\,{4\over l_0^2}\,\bG_{\mu\nu}, \quad
 \bR_{ab}=+\,{4\over l_0^2}\,\bG_{ab}
\ea
for $\mu,\nu \in \{r,0,1,2,3\}$ 
and $a,b\in \{y^1\cdots y^5\}$, 
and have rewritten the expression using the radius $l$ 
of the new $S^5$, which is calculated as 
\ba
 l=\left(1+\frac{2b}{l_0^2}\right)l_0.
\ea 
Note that after the field redefinition, the radius of $S^5$, $l$,  
differs from that of AdS$_5$, $L$, which is expressed as  
\ba
 L \equiv \left(1-{4b\over l^2}\right)l
  =\left(1-{2b\over l_0^2}\right)l_0.
\ea

\resection{Five-Dimensional Effective Action and the Weyl Anomaly}

In this section, we calculate the four-dimensional holographic 
Weyl anomaly from the higher-derivative gravity action \eq{IIB-2}
using the classical solution \eq{sln-2}, 
and show that the resulting anomaly exactly reproduces 
the anomaly of the original gravity theory 
before making the field redefinition. 

To derive the five-dimensional gravity action, we use 
the following strategy. 
First, we assume that the geometry of the ten-dimensional 
spacetime is 
a direct product of a five-dimensional Lorentzian manifold 
$M_5$ and a five-dimensional sphere $S^5$.
Next, we decompose all terms in the action into two parts, 
one of which is expressed by the fields on $M_5$ with metric 
$\widehat{g}_{\mu\nu}$ 
and the other of which is expressed over $S^5$ of radius $l$. 
For example, the scalar curvature $\bR$ in the 
ten-dimensional gravity action becomes $\hR + 20/l^2$. 
(Here $\hR$ is the scalar curvature of $M_5$.) 
However, there appears a problem 
in decomposing the kinetic part of the self-dual 
five-form field strength $F_5$. 
In fact, inserting the classical solution of $F_5$ 
into the action would give a trivial, vanishing result 
due to the self-duality of $F_5$ ($\ast F_5=F_5$).\footnote{%
$\sqrt{-\bG}\left|F_5\right|^2= F_5\wedge\ast F_5=
F_5\wedge F_5=0.$}
To avoid this problem, 
we use the ansatz that $F_5$ has non-zero 
values only for the $S^5$ components, and 
we rescale $F_5$ in the action by the factor $\sqrt{2}$: 
$F_5 \to \sqrt{2}F_5$.
Finally, we integrate over the $S^5$ in the ten-dimensional action 
and obtain the five-dimensional gravity action.

Following this strategy, 
we first calculate the Weyl anomaly of ${\cal N}=4$ $SU(N)$ SYM${}_4$ 
from the action (\ref{IIB-1}). 
Since $\bR=\widehat{R}+20/l_0^2$ and 
$-(1/4)\left|\sqrt{2}F_5\right|^2=-8/l_0^2$, 
we have the five-dimensional action
\ba
 \bS_5 = {\pi^3l_0^5\over 2\kappa_{10}^2g_s^2}\int d^5x
 \sqrt{-\hg}\left({12\over l_0^2}+\hR \right).
\ea
This action actually has an AdS${}_5$ solution 
with radius $l_0$, 
which justifies our ansatz.  
Using the formula (\ref{weyl-2}), we obtain the Weyl anomaly as 
\begin{align} 
 \langle T^i_i \rangle &= \frac{2\pi^3l_0^8}{2\kappa_{10}^2g_s^2}
 \left(-{1\over24}R^2+{1\over8}R_{ij}^2\right) \nn
 &= {N^2\over 4\pi^2}\left(-{1\over24}R^2+{1\over8}R_{ij}^2\right). 
\label{anomaly}
\end{align}
Here we have used $2\kappa_{10}^2 = (2\pi)^7$ and (\ref{radius}).

Next we apply our strategy to the action (\ref{IIB-2}). 
From the solution (\ref{sln-2}), 
we compactify ten-dimensional spacetime on $S^5$ of radius $l$. 
Then, the (dimensionally reduced) five-dimensional action is obtained as  
\begin{align} 
 \widetilde{\bS}_5 = \frac{\pi^3l^5}{2\kappa_{10}^2g_s^2}
 &\left(1+{40a+4b\over l^2} \right) \times \nn
 &\int d^5x \sqrt{-\hg} 
 \left[\left({12\over l^2}-{80a-80b\over l^4}\right)
 +\hR+a\hR^2+b\hR_{\mu\nu}^2\right].
\end{align}
This action has an AdS$_5$ solution with radius 
$\left(1-4b/l^2\right)l$, which is consistent with the 
AdS$_5\times S^5$ solution (\ref{sln-2}). 
From this solution, we can read off the parameters in Eq.\ (\ref{weyl-1}); 
\ba
 {1\over 2\kappa_5^2} = \frac{\pi^3l^5}{2\kappa_{10}^2g_s^2}
 \left(1+{40a+4b\over l^2} \right), \quad 
 L = \left(1-{4b\over l^2}\right)l, \quad c=0.
\ea
Thus the corresponding Weyl anomaly is calculated 
again by using the formula \eq{weyl-2} as
\begin{align}
 \langle T_i^i \rangle &= 
 {2L^3\over 2\kappa_5^2}\left(1-{40a+8b\over l^2}\right)
 \left(-{1\over24}R^2+{1\over8}R_{ij}^2\right) \nn
 &= \frac{2\pi^3l^8}{2\kappa_{10}^2g_s^2}
 \left(1-{16b\over l^2}\right)
 \left(-{1\over24}R^2+{1\over8}R_{ij}^2\right) \nn
 &=\frac{2\pi^3l_0^8}{2\kappa_{10}^2g_s^2}
 \left(-{1\over24}R^2+{1\over8}R_{ij}^2\right) \nn
 &= {N^2\over 4\pi^2}\left(-{1\over24}R^2+{1\over8}R_{ij}^2\right).
\end{align}
This is identical to the result (\ref{anomaly}).

\resection{Conclusion} 

In this paper, we quantitatively checked the validity of the AdS/CFT
correspondence by showing that the holographic RG structure 
is invariant under field redefinitions in Type IIB supergravity. 
In particular, we carried out a redefinition of the ten-dimensional 
metric of the form 
$\bG_{MN}\to\bG_{MN}+\alpha\bR\bG_{MN}+\beta\bR_{MN}$ 
(Eq.\ \eq{redef})
and calculated explicitly the modified Type IIB action. 
We then constructed effective five-dimensional gravity 
when ten-dimensional spacetime is compactified on $S^5$ 
and calculated the holographic Weyl anomaly. 
We showed that the obtained anomaly is identical to 
that of the ${\mathcal{N}}=4$ $SU(N)$ SYM${}_4$ in the large $N$ limit, 
even though the five-dimensional action contains higher-order 
derivative terms. 
This result is consistent with the assertion of 
the AdS/CFT correspondence that 
on-shell fields in the gravity theory are coupled to scaling 
operators of the corresponding CFT at the boundary of the AdS geometry. 
In fact, the theorem of Kamefuchi, O'Raifeartaigh and Salam 
guarantees that a field redefinition does not 
change the on-shell structure of the theory. 

We finally point out that this invariance of the 
holographic Weyl anomaly under a redefinition of the metric 
holds only if there is a simultaneous change 
of the ten-dimensional metric given by \eq{redef}. 
In fact, if we only change the five-dimensional metric 
in the effective five-dimensional action, 
$\widehat{g}_{\mu\nu}\rightarrow \widehat{g}_{\mu\nu}
+\alpha\widehat{R}\,\widehat{g}_{\mu\nu}+\beta\widehat{R}_{\mu\nu}$, 
then the resulting Weyl anomaly 
differs from the field-theoretical anomaly 
in the large $N$ limit. 
However, this is not a contradiction, because if field redefinitions 
are carried out only for five-dimensional components, 
generally the on-shell conditions for a ten-dimensional field theory 
are broken. Thus, there is no reason to expect that 
the AdS/CFT correspondence holds for such redefinitions.

\section*{\large{Acknowledgments}}

The authors would like to thank T.\ Sakai for discussions 
and collaboration in the early stage of this work. 
They also thank M.\ Ninomiya, S.\ Ogushi and T.\ Yokono 
for helpful discussions.


\end{document}